\documentclass[12pt]{article}
\normalbaselines
\global\arraycolsep=2pt

\begin{document}
\bibliographystyle{unsrt}

%\vbox {\vspace{4mm}} %Leave space at the top on the first page.
\begin{center}
{\large \hfill OKHEP-95-06}
\\[2mm]
\end{center}
\begin{center}
{\large \bf Casimir Energy for a Spherical Cavity in a
Dielectric:\\[2mm]Toward a Model for
Sonoluminescence?}\footnote{Invited
Talk at the Third Workshop on Quantum Field Theory
Under the Influence of External Conditions,
Leipzig, September 18--22, 1995.}\\[7mm]
Kimball A. Milton\footnote{e-mail: milton@phyast.nhn.uoknor.edu}\\
{\it Department of Physics and Astronomy\\
The University of Oklahoma, Norman OK 73019, USA}
\\[5mm]
\end{center}

\vspace{2mm}

\begin{abstract}
In the final few years of his life, Julian Schwinger proposed that
the ``dynamical Casimir effect'' might provide the driving force
behind the puzzling phenomenon of sonoluminescence.  Motivated by
that exciting suggestion, I have computed the static Casimir energy
of a spherical cavity in an otherwise uniform material with
dielectric
constant $\epsilon$ and permeability $\mu$.  As expected the result
is divergent;  yet a plausible finite answer is extracted, in the
leading
uniform asymptotic approximation.  That result gives far too small an
energy to account for the large burst of photons seen in
sonoluminescence.
If the divergent result is retained (which is different from that
guessed
by Schwinger), it is of the wrong sign to drive the effect.
Dispersion does not resolve this contradiction.
However, dynamical effects are not yet included.
\end{abstract}
\section{Introduction}
In a series of papers in the last three years of his life, Julian
Schwinger
proposed \cite{js} that the dynamical Casimir effect could provide
the energy
that drives the copious production of photons in the puzzling
phenomenon
of sonoluminescence \cite{sono1,sono2}. In fact, however, he guessed
an approximate
(static) formula for the Casimir energy of a spherical bubble in
water,
based on a general, but incomplete, analysis \cite{rederiv}.  He
apparently
was unaware that I had, at the time I left UCLA, completed the
analysis
of the Casimir force for a dielectric ball \cite{kim}.  It is my
purpose
here to carry out the very straightforward calculation for the
complementary
situation, for a cavity in an infinite dielectric medium.  In fact,
I will consider the general case of spherical region, of radius $a$,
having permittivity $\epsilon'$ and permeability $\mu'$, surrounded
by
an infinite medium of permittivity $\epsilon$ and permeability $\mu$.
% See Fig.~1.

%\begin{figure}
%\begin{picture}(200,300)
%\put(200,100){\circle{200}}
%\put(200,100){\vector(1,1){100}}
%\put(240,50){$a$}
%\put(150,50){$\epsilon'$\quad$\mu'$}
%\put(250,50){$\epsilon$\quad$\mu$}
%\end{picture}
%\caption{Spherical bubble surrounded by infinite medium.}
%\end{figure}

Of course, this calculation is not directly relevant to
sonoluminescence,
which is anything but static.  It is  offered as only a first step,
but it
should give an idea of the orders of magnitude of the energies
involved.
It is an improvement over the crude estimation used in \cite{js}.
Attempts at dynamical calculations exist \cite{sass,eberlein}; but
they are
subject to possibly serious methodological objections.
Sonoluminescence
aside, this calculation is of interest for its own sake, as one of a
relatively few nontrivial Casimir calculations with nonplanar
boundaries
\cite{boyer,balian,mds,dm,bn,fermion,mng,bm}.  It represents a
significant
generalization on the calculation of Brevik and Kolbenstvedt
\cite{brevik},
who consider the same geometry with $\mu\epsilon=\mu'\epsilon'=1$, a
special case, possibly relevant to hadronic physics, in which the
result
is unambiguously finite.

In the next section we review the Green's dyadic formalism we shall
employ, and compute the Green's functions in this case for the TE and
TM
modes.  Then, in Section 3, we compute the force on the shell from
the discontinuity of the stress tensor.  The energy is computed
similarly
in Section 4, and the expected relation between stress and energy is
found.
Estimates in Section 5 show that the result so constructed, even with
physically required subtractions, and including both interior and
exterior
contributions, is divergent, but that if one
supplies a plausible contact term, a finite result (at least in
leading
approximation) follows. (Physically, we expect that the
divergence is regulated by including
dispersion.) Numerical estimates of both the divergent and
finite terms are given in the conclusion, and comparison is made with
the calculations of Schwinger.

\section{Green's Dyadic Formulation}
I follow closely the formulation given in \cite{mds,kim}.  We start
with
Maxwell's equations in rationalized units, with a polarization source
$\bf P$:  (in the following we set $c=\hbar=1$)
\def\vecnab{\mbox{\boldmath{$\nabla$}}}

\begin{eqnarray}
\vecnab\times{\bf H}={\partial\over\partial t}{\bf D}
+{\partial\over\partial t}{\bf P }&,&\qquad \vecnab\cdot{\bf
D}=-\vecnab\cdot
{\bf P},\nonumber\\
-\vecnab\times{\bf E}={\partial\over\partial t}{\bf B} &,&\qquad
\vecnab\cdot{\bf B}=0,
\end{eqnarray}
where, for an homogeneous, isotropic, nondispersive medium
\begin{equation}
{\bf D=\epsilon E,\qquad B=\mu H.}
\end{equation}
\def\GAM{\mbox{\boldmath{$\Gamma$}}}
We define a Green's dyadic $\GAM$ by
\begin{equation}
{\bf E(r},t)=\int (d{\bf r}')\,dt'\, \GAM({\bf r},t;{\bf r}',t')
\cdot{\bf P(r}',t')
\end{equation}
and introduce a Fourier transform in time
\begin{equation}
\GAM({\bf r},t;{\bf r}',t')=\int_{-\infty}^\infty
{d\omega\over2\pi}e^{-i\omega(t-t')}\GAM({\bf r},{\bf r}';\omega),
\end{equation}
\def\PHI{\mbox{\boldmath{$\Phi$}}}
where in the following the $\omega$ argument will be suppressed.
Maxwell's equations then become (which define $\PHI$)
\begin{eqnarray}
\vecnab\times\GAM =i\omega\PHI
&,&\qquad\vecnab\cdot\PHI=0,\nonumber\\
{1\over\mu}\vecnab\times\PHI=-i\omega\epsilon\GAM' &,&\qquad \vecnab
\cdot\GAM'=0,
\label{maxgreen}
\end{eqnarray}
in which $\GAM'=\GAM+{\bf 1}/\epsilon$, where $\bf 1$ includes a
spatial
delta function.  The two solenoidal Green's dyadics given here
satisfy the following second-order equations:
\begin{eqnarray}
(\nabla^2+\omega^2\epsilon\mu)\GAM'&=&-{1\over\epsilon}\vecnab\times
(\vecnab\times{\bf 1}),\\
(\nabla^2+\omega^2\epsilon\mu)\PHI&=&i\omega\mu\vecnab\times
{\bf 1}.
\end{eqnarray}
They can be expanded in vector spherical harmonics
\cite{jackson,stratton}
defined by
\begin{equation}
{\bf X}_{lm}={1\over\sqrt{l(l+1)}}{\bf L}Y_{lm},
\end{equation}
as follows:
\begin{eqnarray}
\GAM'({\bf r,r'})&=&\sum_{lm}\left(f_l(r,{\bf r}'){\bf
X}_{lm}(\Omega)
+{i\over \omega
\epsilon\mu}\vecnab\times g_l(r,{\bf r}'){\bf
X}_{lm}(\Omega)\right),\\
\PHI({\bf r,r'})&=&\sum_{lm}\left(\tilde g_l(r,{\bf r}'){\bf
X}_{lm}(\Omega)
-{i\over \omega}
\vecnab\times \tilde f_l(r,{\bf r}'){\bf X}_{lm}(\Omega)\right).
\end{eqnarray}
When these are substituted in Maxwell's equations (\ref{maxgreen})
we obtain, first,
\begin{equation}
g_l=\tilde g_l,\qquad f_l=\tilde f_l+{1\over\epsilon}{1\over r^2}
\delta(r-r'){\bf X}_{lm}(\Omega'),
\end{equation}
 and then the second-order equations
\begin{eqnarray}
(D_l+\omega^2\mu\epsilon)g_l(r,{\bf r}')&=&i\omega\mu\int d\Omega''\,
{\bf X}^*_{lm}(\Omega'')\cdot\vecnab''\times{\bf 1},\\
(D_l+\omega^2\mu\epsilon)f_l(r,{\bf r}')&=&-{1\over\epsilon}\int
d\Omega''\,
{\bf X}^*_{lm}(\Omega'')\cdot\vecnab''\times(\vecnab''\times{\bf 1}),
\nonumber\\
&=&{1\over\epsilon}D_l{1\over r^2}\delta(r-r'){\bf
X}^*_{lm}(\Omega'),
\end{eqnarray}
where
\begin{equation}
D_l={\partial^2\over\partial r^2}+{2\over r}{\partial\over\partial r}
-{l(l+1)\over r^2}.
\end{equation}

These equations can be solved in terms of Green's functions
satisfying
\begin{equation}
(D_l+\omega^2\epsilon\mu){ F}_l(r,r')=-{1\over r^2}\delta(r-r'),
\end{equation}
which have the form
\begin{equation}
{F}_l(r,r')=\left\{
\begin{array}{ll}
ik'j_l(k'r_<)[h_l(k'r_>)-Aj_l(k'r_>)],&r,r'<a,\\
ikh_l(kr_>)[j_l(kr_<)-Bh_l(kr_<)],&r,r'>a,
\end{array}\right.
\label{form}
\end{equation}
where
\begin{equation}
k=|\omega|\sqrt{\mu\epsilon},\qquad k'=|\omega|\sqrt{\mu'\epsilon'},
\end{equation}
and $h_l=h_l^{(1)}$ is the spherical Hankel function of the first
kind.
Specifically, we have
\begin{eqnarray}
\tilde f_l(r,{\bf r'})&=&\omega^2\mu F_l(r,r'){\bf
X}_{lm}^*(\Omega'),\\
g_l(r,{\bf r'})&=&-i\omega\mu\vecnab'\times G_l(r,r'){\bf
X}_{lm}^*(\Omega'),
\end{eqnarray}
where $F_l$ and $G_l$ are Green's functions of the form (\ref{form})
with the constants $A$ and $B$ determined by the boundary conditions
given below.  Given $F_l$, $G_l$, the fundamental Green's dyadic
is given by
\begin{eqnarray}
\GAM'({\bf r,r}')&=&\sum_{lm}\bigg\{\omega^2\mu F_l(r,r'){\bf
X}_{lm}(\Omega)
{\bf X}_{lm}^*(\Omega')\nonumber\\
&&\quad-{1\over\epsilon}\vecnab\times G_l(r,r'){\bf X}_{lm}(\Omega)
{\bf
%% FOLLOWING LINE CANNOT BE BROKEN BEFORE 70 CHAR
X}_{lm}^*(\Omega')\times{\stackrel{\leftarrow}{\vecnab}}{}'
\nonumber\\
&&\quad+{1\over\epsilon}{1\over r^2}\delta(r-r')\sum_{lm}{\bf
X}_{lm}(\Omega)
{\bf X}_{lm}^*(\Omega')\bigg\}.
\label{gam}
\end{eqnarray}

Now we consider a sphere of radius $a$ centered at the origin,
with properties $\epsilon'$, $\mu'$ in the interior and $\epsilon$,
$\mu$ outside.
Because of the boundary conditions that
\begin{equation}
{\bf E}_\perp, \quad\epsilon E_r,\quad B_r, \quad {1\over\mu}{\bf
B}_\perp
\end{equation}
be continuous at $r=a$, we find for the constants $A$ and $B$ in the
two Green's functions in (\ref{gam})
\begin{eqnarray}
A_F&=&{\sqrt{\epsilon\mu'}\tilde e_l(x')\tilde e_l'(x)-
\sqrt{\epsilon'\mu}\tilde e_l(x)\tilde e_l'(x')\over\Delta_l},\\
B_F&=&{\sqrt{\epsilon\mu'}\tilde s_l(x')\tilde s_l'(x)-
\sqrt{\epsilon'\mu}\tilde s_l(x)\tilde s_l'(x')\over\Delta_l},\\
A_G&=&{\sqrt{\epsilon'\mu}\tilde e_l(x')\tilde e_l'(x)-
\sqrt{\epsilon\mu'}\tilde e_l(x)\tilde
e_l'(x')\over\tilde\Delta_l},\\
B_G&=&{\sqrt{\epsilon'\mu}\tilde s_l(x')\tilde s_l'(x)-
\sqrt{\epsilon\mu'}\tilde s_l(x)\tilde s_l'(x')\over\tilde\Delta_l}.
\end{eqnarray}
Here we have introduced $x=ka$, $x'=k'a$, the Riccati-Bessel
functions
\begin{equation}
\tilde e_l(x)=x h_l(x),\qquad \tilde s_l(x)=xj_l(x),
\end{equation}
and the denominators
\begin{eqnarray}
\Delta_l=\sqrt{\epsilon\mu'}\tilde s_l(x')\tilde e_l'(x)
-\sqrt{\epsilon'\mu}\tilde s_l'(x')\tilde e_l(x),\nonumber\\
\tilde\Delta_l=\sqrt{\epsilon'\mu}\tilde s_l(x')\tilde e_l'(x)
-\sqrt{\epsilon\mu'}\tilde s_l'(x')\tilde e_l(x),
\end{eqnarray}
and have denoted differentiation with respect to the argument by a
prime.

\section{Stress on the sphere}
We can calculate the stress (force per unit area) on the sphere by
computing
the discontinuity of the (radial-radial component) of the stress
tensor:
\begin{equation}
{\cal F}=T_{rr}(a-)-T_{rr}(a+),
\end{equation}
where
\begin{equation}
%% FOLLOWING LINE CANNOT BE BROKEN BEFORE 70 CHAR
T_{rr}={1\over2}\langle[\epsilon(E^2_\perp-E_r^2)
+\mu(H_\perp^2-H_r^2)]\rangle.
\end{equation}
The   vacuum expectation values of the product of field
strengths are given directly by the Green's dyadics computed in
Section 2:
\begin{eqnarray}
i\langle{\bf E(r) E(r')}\rangle&=&\GAM({\bf r,r'}),\\
i\langle{\bf B(r) B(r')}\rangle&=&-{1\over\omega^2}\vecnab\times
\GAM({\bf r,r'})\times{\stackrel{\leftarrow}{\vecnab}}{}',
\end{eqnarray}
where here and in the following
we ignore $\delta$ functions because we are interested in the
{\it limit\/} as $\bf r'\to r$.
It is then rather immediate to find for the stress on the sphere
(the {\it limit\/} $t'\to t$ is assumed)
\begin{eqnarray}
%% FOLLOWING LINE CANNOT BE BROKEN BEFORE 70 CHAR
\cal{F}&=&{1\over2ia^2}\int_{-\infty}^\infty
{d\omega\over2\pi}e^{-i\omega
(t-t')}\sum_{l=1}^\infty{2l+1\over4\pi}\nonumber\\
&&\quad\times\bigg\{(\epsilon'-\epsilon)\left[{k^2\over\epsilon}a^2
%% FOLLOWING LINE CANNOT BE BROKEN BEFORE 70 CHAR
F_l(a+,a+)+\left({l(l+1)\over\epsilon'}
+{1\over\epsilon}{\partial\over
\partial r}r{\partial\over\partial
r'}r'\right)G_l(r,r')\bigg|_{r=r'=a+}
\right]\nonumber\\
&&\quad+(\mu'-\mu)\left[{k^2\over\mu}a^2
G_l(a+,a+)+\left({l(l+1)\over\mu'}+{1\over\mu}{\partial\over
\partial r}r{\partial\over\partial
r'}r'\right)F_l(r,r')\bigg|_{r=r'=a+}
\right]\bigg\}\\
&=&{i\over2a^4}\int_{-\infty}^\infty{dy\over2\pi}e^{-iy\delta}
\sum_{l=1}^\infty{2l+1\over4\pi}x{d\over
dx}\ln\Delta_l\tilde\Delta_l,
\label{unsubstress}
\end{eqnarray}
where $y=\omega a$, $\delta=(t-t')/a$, and
\begin{equation}
\ln\Delta_l\tilde\Delta_l=\ln\left[(\tilde s_l(x')\tilde e_l'(x)-
\tilde s_l'(x')\tilde e_l(x))^2-\xi^2(\tilde s_l(x')\tilde e_l'(x)+
\tilde s_l'(x')\tilde e_l(x))^2\right]+\mbox{constant}.
\end{equation}
Here the parameter $\xi$ is
\begin{equation}
\xi={\sqrt{{\epsilon'\over\epsilon}{\mu\over\mu'}}-1\over
\sqrt{{\epsilon'\over\epsilon}{\mu\over\mu'}}+1}.
\end{equation}

This is not yet the answer.  We must remove the term which would be
present if either medium filled all space (the same was done in the
case of parallel dielectrics \cite{sdm}).  The corresponding
Green's function is
\begin{equation}
F_l^{(0)}=\left\{\begin{array}{ll}
ik'j_l(k'r_<)h_l(k'r_>),&r,r'<a\\
ikj_l(kr_<)h_l(kr_>),&r,r'>a\end{array}\right.
\end{equation}
The resulting stress is
\begin{equation}
{\cal F}^{(0)}={1\over a^3}\int_{-\infty}^\infty{d\omega\over 2\pi}
 e^{-i\omega\tau}
\sum_{l=1}^\infty{2l+1\over4\pi}
\left\{x'[\tilde s_l'(x')\tilde e_l'(x')-\tilde e_l(x')
\tilde s_l''(x')]
-x[\tilde s_l'(x)\tilde e_l'(x)-\tilde e_l(x)
\tilde s_l''(x)]\right\}.
\label{vacstress}
\end{equation}
The final formula for the stress is obtained by subtracting
(\ref{vacstress})
from (\ref{unsubstress}):
\begin{eqnarray}
{\cal
F}&=&-{1\over2a^4}\int_{-\infty}^\infty{dy\over2\pi}e^{iy\delta}
\sum_{l=1}^\infty{2l+1\over4\pi}
\big\{x{d\over dx}\ln\Delta_l\tilde\Delta_l\nonumber\\
&&+2x'[ s_l'(x') e_l'(x')- e_l(x')
 s_l''(x')]
-2x[ s_l'(x) e_l'(x)- e_l(x)
 s_l''(x)]\big\},
\label{stress}
\end{eqnarray}
where we have now performed a Euclidean rotation,
\begin{eqnarray}
y&\to& iy,\quad x\to ix,\quad\tau=t-t'\to i(x_4-x_4')
\quad [\delta=(x_4-x_4')/a],\nonumber\\
\tilde s_l(x)&\to& s_l(x)=\sqrt{\pi x\over 2}I_{l+1/2}(x),\quad
\tilde e_l(x)\to e_l(x)={2\over\pi}\sqrt{{\pi x\over 2}}K_{l+1/2}(x).
\end{eqnarray}

\section{Total energy}

In a similar way we can directly calculate the Casimir energy of the
configuration, starting from the energy density
\begin{equation}
U={\epsilon E^2+\mu H^2\over2}.
\end{equation}
In terms of the Green's dyadic, the total energy is
\def\Tr{\mbox{Tr}\,}
\begin{eqnarray}
E&=&\int(d{\bf r})\,U\nonumber\\
&=&{1\over2i}\int r^2 dr\,d\Omega\left[\epsilon\Tr\GAM({\bf r,r})-
{1\over\omega^2\mu}\Tr\vecnab\times\GAM({\bf
r,r})\times\stackrel{\leftarrow}
{\vecnab}\right]\\
%% FOLLOWING LINE CANNOT BE BROKEN BEFORE 70 CHAR
&=&{1\over2i}\int_{-\infty}^\infty{d\omega\over2\pi}
e^{-i\omega(t-t')}
\sum_{l=1}^\infty(2l+1)\int_0^\infty r^2\,dr\nonumber\\
&&\times\left\{2k^2[F_l(r,r)+G_l(r,r)]
+{1\over r^2}{\partial\over\partial r}r{\partial\over\partial r'}r'
[F_l+G_l](r,r')\big|_{r'=r}\right\},
\label{fullen}
\end{eqnarray}
where there is no explicit appearance of $\epsilon$ or $\mu$.
(However, the value of $k$ depends on which medium we are in.)
As in \cite{mds} we can easily show that the total derivative term
integrates
to zero. We are left with
\begin{equation}
E={1\over2i}\int_{-\infty}^\infty{d\omega\over2\pi}e^{-i\omega\tau}
\sum_{l=1}^\infty(2l+1)\int_0^\infty r^2\,dr\,
2k^2[F_l(r,r)+G_l(r,r)].
\label{en}
\end{equation}
However, again we should subtract off that contribution which the
formalism
would give if either medium filled all space.  That means we should
replace $F_l$ and $G_l$ by
\begin{equation}
\tilde F_l, \tilde G_l=\left\{\begin{array}{ll}
-ik'A_{F,G}j_l(k'r)j_l(k'r'),&r,r'<a\\
-ikB_{F,G}h_l(kr)h_l(kr'),&r,r'>a
\end{array}\right.
\label{rep}
\end{equation}
so then (\ref{en}) says
\begin{equation}
E=-\sum_{l=1}^\infty(2l+1)\int_{-\infty}^\infty{d\omega\over
2\pi}e^{-i\omega\tau}
\left\{\int_0^a r^2dr\,k^{\prime 3}(A_F+A_G)j_l^2(k'r)
+\int_a^\infty r^2dr\,k^3(B_F+B_G)h_l^2(kr)\right\}.
\end{equation}
The radial integrals may be done by using the following indefinite
integral for any spherical Bessel function $j_l$:
\begin{equation}
\int dx\,x^2j_l^2(x)={x\over2}[((xj_l)')^2-j_l(xj_l)'-xj_l(xj_l)''].
\end{equation}
But we must remember to add the contribution of the total derivative
term
in (\ref{fullen}) which no longer vanishes when the replacement
(\ref{rep})
is made.  The result is precisely that expected from the stress
(\ref{stress}),
\begin{equation}
E=4\pi a^3{\cal F},\qquad {\cal F}={1\over4\pi a^2}
\left(-{\partial\over\partial a}
\right)E,
\end{equation}
where the derivative is the naive one, that is,  the cutoff
$\delta$ has no effect on the derivative.

\section{Asymptotic analysis and numerical results}

The result for the stress (\ref{stress}) is an immediate
generalization
of that given in \cite{kim}, and therefore, the asymptotic analysis
given there can be applied nearly unchanged.  The result for the
energy
is new, and seems not to have been recognized earlier.

We first remark on the special case
$\sqrt{\epsilon\mu}=\sqrt{\epsilon'\mu'}$.
Then $x=x'$ and the energy reduces to
\begin{equation}
E=-{1\over4\pi a}\int_{-\infty}^\infty dy\,
e^{iy\delta}\sum_{l=1}^\infty
(2l+1)x{d\over dx}\ln[1-\xi^2((s_le_l)')^2],
\end{equation}
where
\begin{equation}
\xi={\mu-\mu'\over\mu+\mu'}.
\label{emu}
\end{equation}
If $\xi=1$ we recover the case of a perfectly conducting spherical
shell, treated in \cite{mds}, for which $E$ is finite.  In fact
(\ref{emu})
is finite for all $\xi$, and if we use the leading uniform asymptotic
approximation for the Bessel functions we obtain
\begin{equation}
U\sim {3\over64a}\xi^2.
\end{equation}
Further analysis of this special case is given by Brevik and
Kolbenstvedt
\cite{brevik}.

In general, using the uniform asymptotic behavior, with $x=\nu z$,
$\nu=l+1/2$, and, for simplicity looking at the large $z$ behavior,
we have
\begin{equation}
E\sim-{1\over 2\pi
a}{1\over\sqrt{\epsilon\mu}}\sum_{l}\nu^2\int_{-\infty}
^\infty dz\,e^{iz\nu\delta/\sqrt{\epsilon\mu}}
z{d\over
dz}\ln\left[1+{1\over16z^4}\left({\epsilon\mu\over\epsilon'\mu'}
-1\right)^2(1-\xi^2)\right],
\end{equation}
which exhibits a cubic divergence as $\delta\to0$.  To be more
explicit,
let us content ourselves with with the case when $\epsilon-1$,
$\epsilon'-1$
are both small and $\mu=\mu'=1$.  Then, the leading $\nu$ term is
\begin{eqnarray}
E&\sim&-{(\epsilon'-\epsilon)^2\over16\pi
a}\sum_{l=1}^\infty\nu^2{1\over2}
\int_{-\infty}^\infty dz\,e^{i\nu z\delta}z{d\over
dz}{1\over(1+z^2)^2}
\nonumber\\
&=&-{(\epsilon'-\epsilon)^2\over64
a}\left({16\over\delta^3}+{1\over4}
\right)\to-{(\epsilon'-\epsilon)^2\over256 a}.
\label{smalle}
\end{eqnarray}
Here, the last arguable step is made plausible by noting that since
$\delta=\tau/a$ the divergent term represents a contribution to the
surface tension on the bubble, which should be cancelled by a
suitably
chosen counter term (contact term). This argument is given somewhat
more weight by the discussion in \cite{toward}.
In essence, justification is provided there for the use of
zeta-function regularization, which directly gives the finite part
here:
\begin{equation}
E\sim{(\epsilon'-\epsilon)^2\over32\pi a}\sum_{l=1}^\infty\nu^2
{\pi\over2}={(\epsilon'-\epsilon)^2\over64 a}\left(-{1\over4}\right),
\end{equation}
because $\sum_{l=0}^\infty\nu^s=(2^{-s}-1)\zeta(-s)$ vanishes at
$s=2$.

Alternatively, one could argue that
dispersion should be included \cite{brevein,candelas,bss},
crudely modelled by
\begin{equation}
\epsilon(\omega)-1={\epsilon_0-1\over1-{\omega^2/\omega_0^2}}.
\end{equation}
If this rendered the expression for the stress finite [we consider
the stress, not the energy, for it is not necessary to consider the
dispersive factor $d(\omega\epsilon(\omega))/d\omega$ there], we
could
drop the cutoff $\delta$ and the sign of the force would be positive:
(at last, we set $\epsilon'=1$)
\begin{equation}
{\cal F}\sim+{(\epsilon_0-1)^2\over128\pi^2a^4}
\sum_{l=1}^\infty\nu^2\int_{-\infty}^\infty
dz\,{1\over(1+z^2)^2}{1\over(1+z^2/z_0^2)^2},
\label{dispforce}
\end{equation}
where $z_0=\omega_0a/\nu$.  As $\nu\to\infty$, $z_0\to0$, and
the integral here approaches $\pi z_0/2$, and so
\begin{equation}
{\cal F}\sim{(\epsilon_0-1)^2\over256\pi
a^3}\omega_0\sum_{l=1}^{\nu_c}
\nu\sim{(\epsilon_0-1)^2\over512\pi a}\omega_0^3,
\label{dispforce2}
\end{equation}
if we take as the cutoff\footnote{Inconsistently, for then
$z_0\sim1$.  If $z_0=1$ in (\ref{dispforce}), however, the same
angular momentum cutoff gives $5/12$ of the value in
(\ref{dispforce2}).}
of the angular momentum sum $\nu_c\sim\omega_0a$.
The corresponding energy is obtained by integrating $-4\pi a^2{\cal
F}$,
\begin{equation}
E\sim-{(\epsilon_0-1)^2\over256}\omega_0^3a^2,
\label{incldisp}
\end{equation}
which is of the form of (\ref{smalle}) with $1/\delta\to\omega_0/4$.

\section{Conclusions}

So finally, what can we say about sonoluminescence?  To calibrate our
remarks, let us recall (a simplified version of) the argument of
Schwinger
\cite{js}.  On the basis of a provocative but
incomplete analysis he argued that
a bubble ($\epsilon'=1$) in water ($\epsilon\approx (4/3)^2$)
possessed a positive Casimir energy\footnote{Note, for small
$\epsilon-1$, Schwinger's result goes like $(\epsilon-1)$,
indicating that he had not removed the ``vacuum'' contribution
corresponding to (\ref{vacstress}).  This is the essential physical
reason for the discrepancy between his results and mine.}
\begin{equation}
E_c\sim{4\pi a^3\over3}\int{(d{\bf k})\over(2\pi)^3}{1\over2}k\left(
1-{1\over\sqrt{\epsilon}}\right)\sim{a^3
K^4\over12\pi}\left(1-{1\over
\sqrt{\epsilon}}\right),
\end{equation}
where $K$ is a wavenumber cutoff.  Putting in his estimate,
$a\sim 4\times 10^{-3}$cm, $K\sim2\times 10^{5}$cm$^{-1}$ (in the
UV),
we find a large Casimir energy, $E_c\sim 13$MeV, and something like
3 million photons would be liberated if the bubble collapsed.

What does our full (albeit static) calculation say?  If we believe
the subtracted result, the last form in (\ref{smalle}), and say that
the bubble collapses from an initial radius $a_i=4\times 10^{-3}$cm
to a final radius $a_f=4\times10^{-4}$cm, we find that the change in
the Casimir energy is $\Delta U\sim+10^{-4}$eV.  This is far to small
to account for the observed emission.

On the other hand, perhaps we should retain the divergent result, and
put in reasonable cutoffs.  If we do so, we have
\begin{equation}
E=-{(\epsilon-1)^2\over 4}a^2K^3\sim-4\times10^5\mbox{eV},
\end{equation}
perhaps of acceptable magnitude, but of the {\it wrong\/} sign.
The same conclusion follows if one uses dispersion, as
(\ref{incldisp})
shows.

So we are unable to see how the Casimir effect could possibly
supply energy relevant to the copious emission of light seen in
sonoluminescence.  Of course, dynamical effects could change this
conclusion, but elementary arguments suggest that this is impossible
unless ultrarelativistic velocities are achieved.  (See also
\cite{eberlein}.)
Yet the subject of vacuum energy is sufficiently subtle that
surprises
could be in store.
A more complete analysis will be provided elsewhere.
\section*{Acknowledgments}
I thank the US Department of Energy for partial financial support of
this
research.  I am grateful to M. Bordag and the other organizers
 for inviting me to participate in this very interesting
Workshop and for providing such gracious hospitality.
  I am happy to acknowledge
 useful conversations with J. Ng, I. Brevik, and C. Eberlein.
I dedicate this paper to the memory of Julian Schwinger, who taught
me
so much, and first stirred my interest in this subject and then
revived that
interest, with his remarkable suggestion that the Casimir effect is
behind
sonoluminescence.  Would that it be so!


\begin{thebibliography}{99}

\bibitem{js} J. Schwinger, Proc.\ Natl.\ Acad.\ Sci.\ USA {\bf 90},
958, 2105, 4505, 7285 (1993); {\bf 91}, 6473 (1994).

\bibitem{sono1} H. Frenzel and H. Schultes, Z. Phys.\ Chem., Abt.\ B
{\bf 27}, 421 (1934).

\bibitem{sono2} D. F. Gaitan, L. A. Crum, C. C. Church, and R. A.
Roy,
J. Acoust.\ Soc.\ Am.\ {\bf 91}, 3166 (1992);
B. P. Barber and S. J. Putterman, Nature {\bf 352}, 318 (1991); B. P.
Barber, R. Hiller, K. Arisaka, H. Fetterman, and S. J. Putterman,
J. Acoust.\ Soc.\ Am.\ {\bf 91}, 3061 (1992); R. G. Holt, D. F.
Gaitan,
A. A. Atchley, and J. Holzfuss, Phys.\ Rev.\ Lett.\ {\bf 72}, 1376
(1994); R. Hiller, S. J. Putterman, and B. P. Barber, Phys.\ Rev.\
Lett.\
{\bf 69}, 1182 (1992); B. P. Barber and S. J. Putterman, Phys.\ Rev.\
Lett.\ {\bf 69}, 3839 (1992).

\bibitem{rederiv} J. Schwinger, Lett.\ Math.\ Phys.\ {\bf 24}, 59,
227 (1992);
Proc.\ Natl.\ Acad.\ Sci.\ USA {\bf 89}, 4091, 11118 (1992).

\bibitem{kim} K. A. Milton, Ann.\ Phys.\ (N.Y.) {\bf 127}, 49 (1980).

\bibitem{sass} E. Sassaroli, Y. N. Srivastava, and A. Widom,
Phys.\ Rev.\ D {\bf 50}, 1027 (1993).

\bibitem{eberlein} C. Eberlein, Illinois preprints P-95-05-037
(quant-ph/9506023), P-95-06-039 (quant-ph/9506024).

\bibitem{boyer} T. H. Boyer, Phys.\ Rev.\ {\bf 174}, 1764 (1968).

\bibitem{balian} R. Balian and B. Duplantier, Ann.\ Phys.\ (N.Y.)
{\bf 112},
165 (1978).

\bibitem{mds} K. A. Milton, L. L. DeRaad, Jr., and J. Schwinger,
Ann.\ Phys.\ (N.Y.) {\bf 115}, 388 (1978).

\bibitem{dm} L. L. DeRaad, Jr. and K. A. Milton, Ann.\ Phys.\ (N.Y.)
{\bf 136}, 229 (1981).

\bibitem{bn} I. Brevik and G. H. Nyland, Ann. Phys.\ (N.Y.)
{\bf 230}, 321 (1994).

\bibitem{fermion} K. A. Milton, Ann.\ Phys.\ (N.Y.) {\bf 150}, 432
(1983).

\bibitem{mng} K. A. Milton and Y. J. Ng, Phys.\ Rev.\ D {\bf 46}, 842
(1992).

\bibitem{bm} C. M. Bender and K. A. Milton, Phys.\ Rev.\ D {\bf 50},
6547 (1994).

\bibitem{brevik} I. Brevik and H. Kolbenstvedt, Phys.\ Rev.\ D {\bf
25},
1731 (1982); Phys.\ Rev.\ D {\bf 26}, 1490 (1983) (E);
 Ann.\ Phys.\ (N.Y.) {\bf 143}, 179 (1982); {\bf 149}, 237 (1983).


\bibitem{jackson} J. D. Jackson, {\it Classical Electrodynamics}, 2nd
ed.
(Wiley, New York, 1975), p.~746 ff.

\bibitem{stratton} J. A. Stratton, {\it Electromagnetic Theory}
(McGraw-Hill, New York, 1941), Chapter 7.

\bibitem{sdm} J. Schwinger, L. L. DeRaad, Jr, and K. A. Milton,
Ann.\ Phys.\ (N.Y.) {\bf 115}, 1 (1978).

\bibitem{toward} K. A. Milton, Phys.\ Rev.\ D {\bf 27}, 439 (1983).

\bibitem{brevein} I. Brevik and G. Einevoll, Phys.\ Rev.\ D {\bf 37},
2977
(1988); I. Brevik and I. Clausen, Phys.\ Rev.\ D {\bf 39}, 603
(1989).

\bibitem{candelas} P. Candelas, Ann.\ Phys.\ (N.Y.) {\bf 143}, 241
(1982);
{\bf 167}, 257 (1986).

\bibitem{bss} I. Brevik, H. Skurdal, and R. Sollie, J. Phys.\ A
{\bf 27}, 6853 (1994).

\end{thebibliography}
\end{document}